\documentstyle[11pt,paspconf,epsf]{article}

\begin{document}

\title{Comments On The Intrinsic X-ray/UV Absorbers In AGN}
\author{Smita Mathur}
\affil{Harvard-Smithsonian Center for Astrophysics, 60 Garden St.,
Cambridge, MA 02138}

\begin{abstract}

 This talk discusses the unified X-ray/UV absorption
models in AGN. I will first review  the models and discuss
how they offer an unique opportunity to probe the near nuclear
environment of AGNs. Recently, however,
the validity of these models has been questioned by a number of
authors.  I will discuss these papers and
argue that the data/ models presented in them do NOT contradict the
unified X-ray/UV picture. I will conclude by presenting our new HST
spectrum of a
X-ray warm absorber.

\end{abstract}


\section{Introduction}

 About 10\% of AGN show `associated absorption' in their UV
spectra. While this absorbing material must
be associated with the active
nucleus, it does not appear to be due to any known component of an AGN
and  there is no accepted model for it (Foltz {\it et~al}~1988).

 Similarly, many AGNs, mainly
Seyfert galaxies, exhibit strong low energy X-ray cut-offs due to
neutral (``cold'') material in their nuclei, also with no accepted
identification. Recent observations of X-ray ionized (``warm'')
absorbers showed the existence of high column density, highly ionized
material in the near nuclear region. Due to their apparently different
physical properties, the possibility of linking any of these three
types of absorbers has not seemed promising. Only a limited understanding of
the physical conditions in the absorbing material is obtained through
UV or X-ray studies alone since the ionization structure of the
absorber is not known.

\begin{quote}
\centering { Absorption Systems in Quasar Spectra}
\end{quote}
\vspace*{-0.35in}
\begin{table}[h]
\begin{center}
\begin{tabular}{|lll|}
\hline
UV &X-ray (Cold) & X-ray (Warm)\\
&&\\
\hline
\hline
Low N$_H$ ($< 10^{20}$)& High N$_H$ ($> 10^{21}$) & High N$_H$ ($>
10^{21}$) \\
&&\\
Ionized & Neutral & Ionized \\
Low U ($ 10^{-3}$)& & High U ($>$1)\\
 & & \\
\hline
\end{tabular}
\end{center}
\end{table}

 This situation changed with quasi-simultaneous observations of the
quasar 3C351 with ROSAT and HST. The X-ray spectrum revealed the
presence of a warm absorber showing K-edge due to OVII or OVIII (Fiore
et al. 1993). The
UV spectrum showed strong, associated, high ionization absorption
lines including OVI $\lambda \lambda 1031, 1037$ (Bahcall {\it
et~al}~1993). Through detailed
modeling we found an excellent match between the X-ray and UV
absorber properties (Mathur {\it et~al}~1994 ). {\it The combination of
both X-ray and
UV datasets allows strong constraints to be placed upon the physical
conditions of the absorber}. In 3C351 we find that the absorber has
well determined properties which describe {\it a component of nuclear
material not previously recognized:} highly ionized, outflowing, low
density, high column density material situated outside the broad
emission line region.

 Since the discovery of a unified X-ray/UV absorber in an X-ray
quiet quasar 3C351, we have found that the X-ray and UV absorber are
also one and the same in other, quite different, AGN: a `red' quasar
3C212 (Mathur 1994) and a variable Seyfert galaxy NGC5548 (Mathur {\it
et~al}~1995 ). We
developed a model for an evolving X-ray/UV absorber in NGC3516 which
explains the previous presence and present disappearance of its broad
absorption lines  (Mathur {\it et~al}~1997). It makes definite
predictions about the
evolution of its X-ray warm absorber which can be tested with future
X-ray missions. A unified X-ray/UV absorber is also found in NGC 3783
(Shields, J. these proceedings). M. Crenshaw and collaborators (these
proceedings)  have
also found the common occurrence of UV and X-ray absorption in Seyfert
1 galaxies with high signal-to-noise HST spectra.  These results open up the
possibility that this
unified picture is quite general. It suggests that a wide range of
associated absorbers may in fact be related through a continuum of
properties like the column density, ionization parameter, distance
from the continuum source, and the outflow velocity. These results may
be extended to the broad absorption line quasars (BALQSOs). The
prototype BALQSO PHL5200, the only BALQSO with a X-ray spectrum, shows
strong X-ray absorption (Mathur, Elvis \& Singh 1995). (See Paul
Green's article in these proceedings for X-ray properties of BALQSOs).

 Understanding of the physical properties of the absorber allowed us
 to pose astrophysical questions about the geometry and kinematics of the
absorbing outflow. We found that the outflow is likely to be edge-on
with the kinetic energy a large fraction of the bolometric luminosity
of the AGN (Mathur {\it et~al}~1995). This is similar to the large
out-flow energy discussed by J. Miller yesterday.

\section{The Debate}

 Clearly, the unified picture of X-ray/UV absorbers offers an unique
opportunity to probe the nuclear environment of AGNs. It was greeted
by the AGN community with both enthusiasm and skepticism. Enthusiasm
for obvious reasons, that it offered something new, interesting and
useful. Skepticism because the
X-ray and UV absorbers don't {\it have} to be one and the same: (1) A small
N$_H$, small U solution
still exists for UV absorbers, so they need not be the same as X-ray
absorbers; (2) there are theoretical models of X-ray absorbers (Krolik
\& Kriss (K\&K) 1995 ) with
much larger N$_H$ and larger U which do not produce UV absorption
lines.

 The K\&K models for X-ray warm absorbers can be easily
distinguished from our X/UV models by their Fe-K edge. The  K\&K
models predict the presence of an Fe-K edge in addition to the
OVII/OVIII edges while the X/UV models do not. I would like to point
out that an Fe-K edge was not detected in ASCA observations of NGC5548
and NGC3516. It was detected in NGC3783, however the warm absorber
model fit to the OVII/OVIII edges could not simultaneously fit the
Fe-K edge (George {\it et~al}~1995). This implies a different origin for
the gas producing Fe-K edge from that producing the OVII/OVIII warm
absorber. This might also be the case in other AGNs showing Fe-K edges
in their {\it Ginga} spectra.

\subsection{Further Debate}

 The uniqueness of the combined X-ray-UV absorber has been questioned
recently. I will  discuss the relevant  papers in turn.\\

\noindent
1. A detailed paper by  Netzer (1996) discusses X-ray lines in AGN \&
photoionized gas. The models presented in this paper span a range of
parameter space in density, column density and ionization
parameter. The author claims that the X-ray warm absorber model for
NGC 5548 {\it cannot} reproduce the strengths of the observed UV
absorption lines whereas Mathur et al. (1995) present a detailed
analysis demonstrating their consistency. The reason for this apparent
contradiction is the different input continuum shape.
The input continuum assumed by Netzer leads to
N(CIV) $<$ N(NV) and the resulting disagreement with the
observations. We, on the other hand,  used the {\it observed}
continuum for NGC5548 which
 contains an additional  strong soft excess characterized by a black-body
spectrum of temperature 150,000K. For this continuum N(CIV) $>$ N(NV)
as observed. This illustrates not only the danger of comparing results
using differing
assumptions but also the importance of using the observed data. It may
also offer a way of determining the unobservable EUV
continuum.\\

\noindent
2. NGC3516: The Seyfert galaxy NGC3516 contains the strongest UV
associated absorption line system. It has both broad and narrow lines
of both high and low ionization. It also contains a X-ray warm
absorber. Clearly, it has a complex, multicomponent absorption system.
Kriss {\it et~al}~ (1996a,b) present  simultaneous
HUT and ASCA observations of NGC 3516. They conclude that a unified
X-ray/UV absorber model is not applicable for NGC 3516.

We note, however, that the X/UV model generally associates broad UV
absorption with X-ray absorption. In NGC3516, the broad lines
{\it disappeared} since $\sim$1992 (Koratkar {\it et~al}~ 1996).
Kriss {\it et~al}~ compared the narrow UV lines with the
X-ray warm absorber and found the two to be incompatible, which is
 not surprising. Based on ROSAT data, we found that the
properties of the warm absorber are in fact consistent with the
present non-detection of the broad high ionization lines (Mathur,
Wilkes \& Aldcroft 1997). We developed an evolving model in which the
ionization parameter of the absorber increases due to a decrease in
density as it expands while out-flowing. This scenario is consistent
with the presence of broad lines in the past and their subsequent
disappearance. This X-ray/UV absorber model makes definite predictions
about the future (e.g. the OVIII edge would get stronger than the OVII
edge) which can be tested with missions like AXAF and XMM. \\

\noindent
3. Other Cases: Hamann (1997) has found some cases in which the
overall level of ionization is not consistent with single zone models
of the UV line  and X-ray warm absorbers. In the z$_a
\approx$z$_e$ systems discussed in this paper low as well as high
ionization lines are present. Clearly, a range of ionizations is
required to produce these lines. The X-ray warm absorber typically has
a high ionization parameter defined by its OVII/OVIII ratio. It cannot
contribute to the low ionization lines. This is entirely as expected
from the X/UV models in which the X-ray warm absorber produces the
high ionization absorption lines in the UV. Note also that none of the
z$_a\approx$z$_e$ systems in Hamann (1997) has a known X-ray warm absorber. A
direct comparison of the X-ray and UV absorbers thus cannot be made.

\section{PG1114+445: New HST Spectra}

 The ROSAT  spectrum of PG1114+445 shows that it contains a
X-ray warm absorber (Laor {\it et al.}~1994). Based on the unified X-ray/UV
absorber picture,
we expected it to show high ionization UV absorption lines. To test this
prediction we observed PG1114+445 with HST FOS. The UV
spectrum clearly shows the absorption lines of CIV, NV and
Ly${\alpha}$ at 1770\AA, 1418\AA\ and 1389\AA\ respectively (Figure
1). The low
ionization MgII absorption line was not detected,
as expected (spectrum not shown here).

\begin{figure}
\caption{The HST FOS Spectrum Of PG1114+445}
\vspace*{0.2in}
\plotone{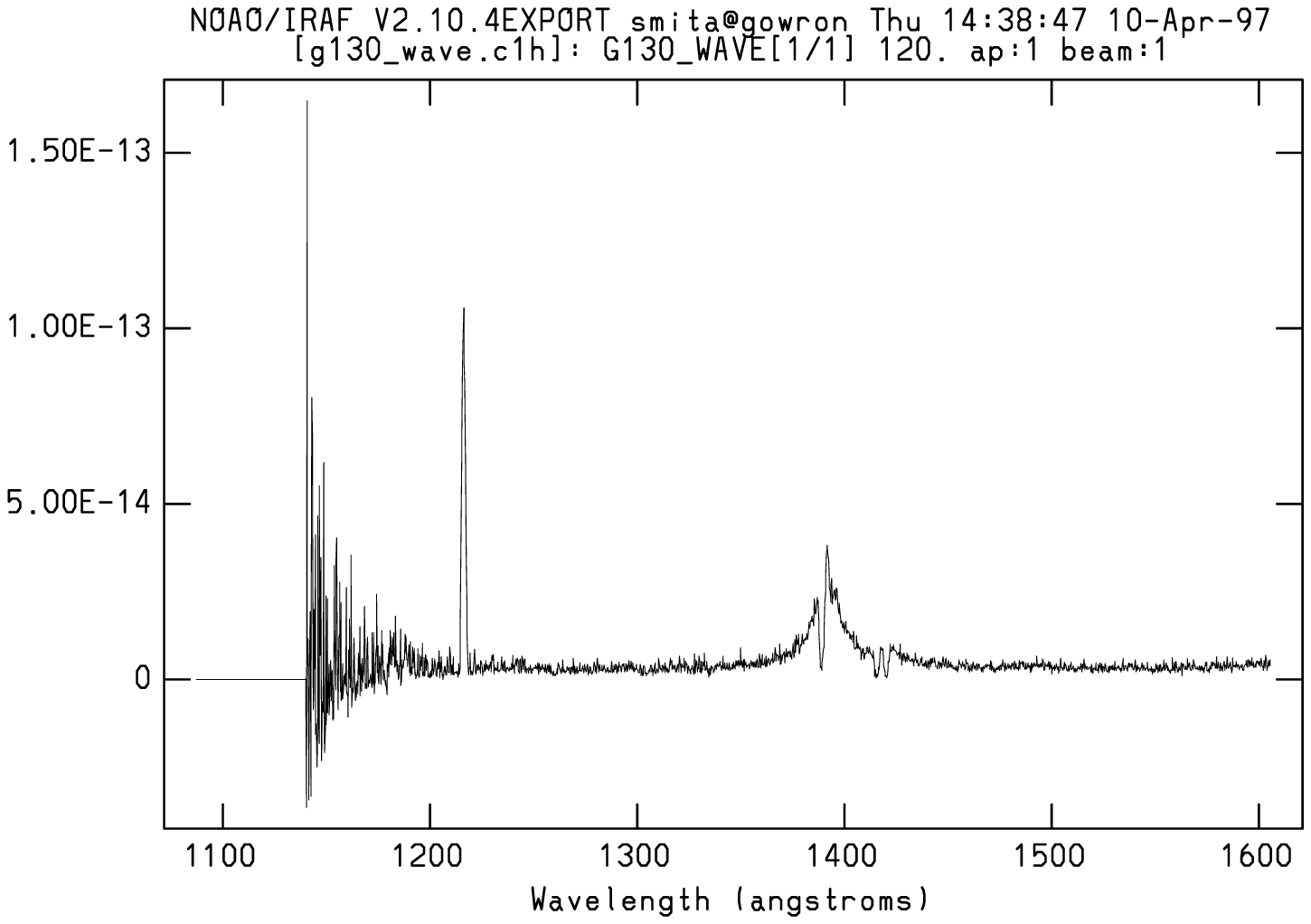}
\vspace*{0.5in}
\plotone{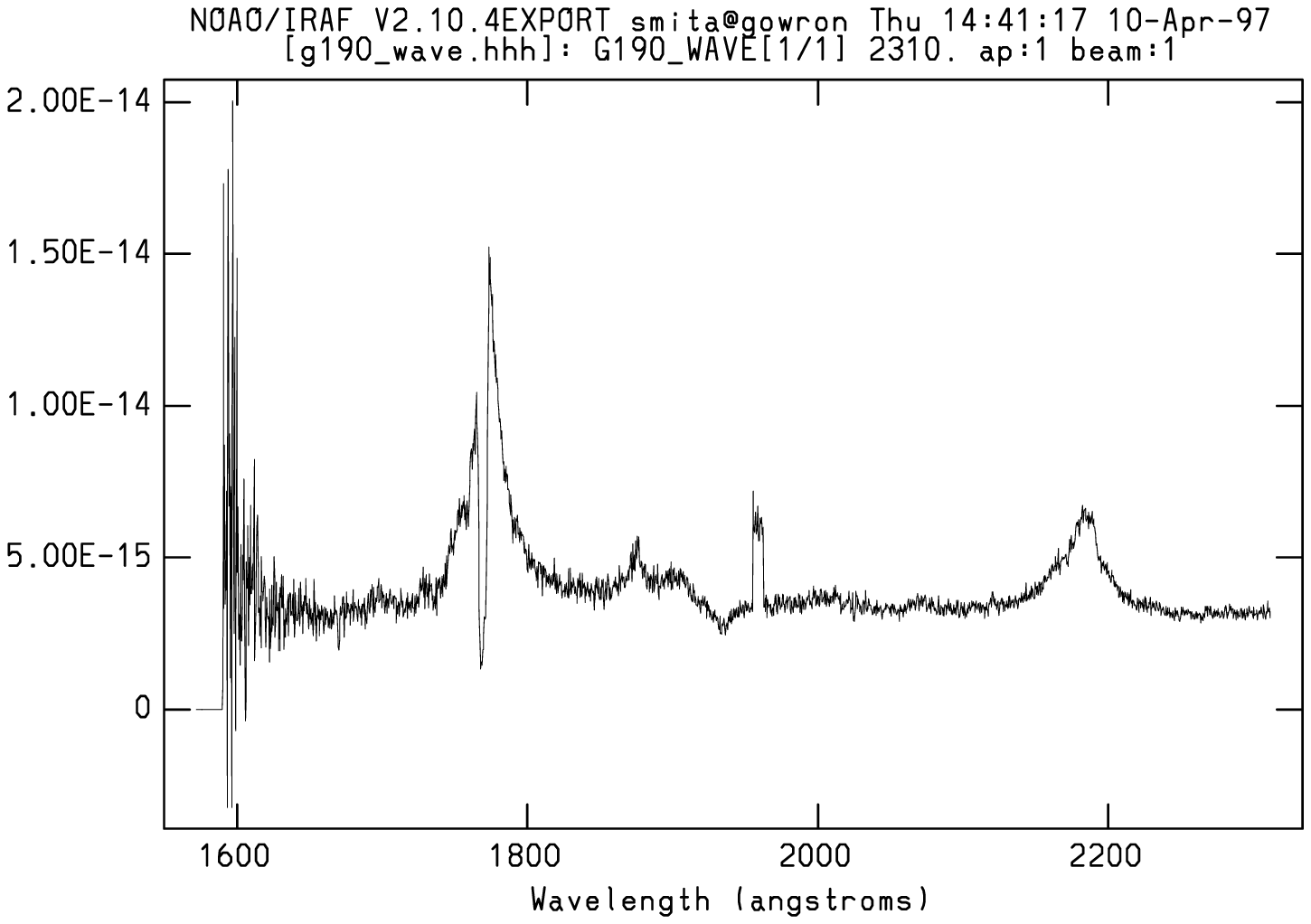}
\end{figure}

Can this be just a chance coincidence?
About 10\% of Seyfert galaxies have associated absorption lines in
their UV spectra (Ulrich 1988). So, there is a 10\% chance of
finding them in any randomly selected Seyfert galaxy. Strong
associated absorption also appears to arise predominantly in steep spectrum,
radio-loud objects (Foltz {\it et al.}~1988)
rather than in radio-quiet quasars. In fact, there has been some
evidence of dichotomy between the occurrence of associated absorption
in radio-loud quasars and broad absorption lines in radio-quiet
quasars (BALQSOs) (Foltz {\it et al.}~1988). It is thus highly unlikely
that the existence of associated absorption lines in radio-quiet
 PG1114+445  is  a chance coincidence. This
observation suggests that the X-ray and UV absorbers
are physically related. A few more such tests could be conclusive.

\section{Conclusions}

 Present evidence argues strongly in favor of the unified picture of
the X/UV absorbers with the X-ray and UV absorbers  physically
related, if not identical. The X-ray warm absorbers, with their large
column densities, would typically contribute strongly to the
broad, high ionization absorption lines seen in the UV. As discussed
in $\S$2, the UV line systems may not {\it require} continuum
absorption in X-rays, but the X-ray absorbers {\it must} produce the
high ionization absorption lines.

 Even though the unified model of X-ray and UV absorbers is a good
description of the observations, it may not be adequate to explain complex
systems in its present
single-zone, photoionization equilibrium form. Multi-zone, multi-parameter
models will be required when the absorption systems show multiple
components (e.g NGC3516, objects in Hamann 1997). In some highly
variable objects (e.g. NGC 4051),
non-equilibrium models might be necessary (Nicastro {\it
et~al.}~1997). The X/UV models should evolve as better
quality of data becomes available. These models, however, provide an unique
opportunity and a good
starting point for understanding the associated absorption systems and so
probe the near nuclear environment of AGN.

\acknowledgments

 I would like to thank the organizers  for a wonderful and stimulating
workshop. The work presented here is supported by grants
NAG5-3249 (LTSA) and  GO-06484.01-95A (from STScI).

%\begin{question}{}

%\end{question}
%\begin{answer}

%\end{answer}

\end{document}